\documentclass[twocolumn,prl,sort&compress,floatfix]{revtex4-1}

\usepackage{amssymb}
\usepackage{amsmath}
\usepackage{mathrsfs}
\usepackage{enumitem}
\usepackage{cleveref}
\usepackage{bm}
\usepackage{tabularx}
\usepackage[dvipsnames]{xcolor}
\usepackage[sort&compress]{natbib}
\usepackage{graphicx}
\usepackage{color}
\usepackage{xr}
\usepackage{soul}
\usepackage{lineno}
\usepackage{lipsum} 
\usepackage{nonfloat}

\usepackage{mathtools}
\begin{document}

\title{Nucleoid clustering drives stepwise expansion and segregation of replicating bacterial chromosomes}

\author{G. Forte$^1$, E. Orlandini$^2$, D. Marenduzzo$^1$, }
\affiliation{$^1$ SUPA, School of Physics and Astronomy, University of Edinburgh, Peter Guthrie Tait Road, Edinburgh, EH9 3FD, UK\\  
$^2$ Dipartimento di Fisica e Astronomia and Sezione INFN, Universit\'a di Padova, Via Marzolo 8, Padova 35131, Italy.}


\begin{abstract}

Bacterial chromosome replication occurs in the absence of a canonical spindle apparatus; yet it reliably produces organised and segregated genomes. While both passive and active mechanisms have been investigated, DNA replication itself is a non-equilibrium process that continuously generates new genetic material and reorganizes the nucleoid. Here, we investigate how replication-driven dynamics, combined with nucleoid-associated protein (NAP) interactions, shape spatiotemporal chromosome organisation using a three-dimensional polymer model that explicitly simulates DNA synthesis. We show that NAP-mediated interactions induce dynamic clustering of DNA, generating density fluctuations in the nucleoid. When coupled to replication, these clusters undergo cycles of stress buildup and release that produce stepwise expansion dynamics consistent with experimental observations. Chromosome segregation occurs naturally in this regime, but only within a finite range of interaction strengths: weak interactions fail to structure the nucleoid, whereas strong interactions hinder replication progression. Within this optimal balance, replication also promotes the spontaneous formation of replication factories. Our results demonstrate that bacterial chromosome organisation can be understood as a non-equilibrium system in which the interplay between replication forces and protein-mediated interactions generates nucleoid mechanics, dynamics, and segregation.

\end{abstract}

\maketitle

The bacterial chromosome is a highly dynamic polymer that is continuously driven out of equilibrium by cellular processes such as DNA replication, transcription, and loop extrusion~\cite{Wang2013,Badrinarayanan2015}. During the cell cycle, these processes reorganise the genome while simultaneously generating new DNA and altering its topology, placing the nucleoid in an intrinsically non-equilibrium state. How such an active system achieves robust chromosome organisation and segregation in the absence of a canonical spindle apparatus remains a central open question in bacterial cell biology~\cite{Wang2013,Badrinarayanan2015,Teleman1998,Jun2006,Harju2024,Brahmachari2026}.

Fluorescence microscopy experiments have revealed that the nucleoid also exhibits some striking spatiotemporal dynamics, including transient density waves during interphase (G1) and stepwise expansion during DNA replication~\cite{Fisher2013}. These observations point to a highly dynamic internal organisation, in which the nucleoid undergoes subsequent steps of intermittent compaction and release. Despite extensive experimental characterisation, the physical mechanisms underlying these behaviors, and their biological function, remain poorly understood. In particular, it is unclear how the interplay between passive forces, such as confinement and polymer entropy~\cite{Jun2006}, and active processes associated with replication and chromosome organisation~\cite{Forte2024,Forte2026} gives rise to both the observed dynamics and the reliable segregation of sister chromosomes.

Entropic models have shown that confinement can drive demixing of replicated polymers, leading to spontaneous segregation at equilibrium~\cite{Jun2006,Jun2010,Mitra2022}. However, these mechanisms primarily operate once replication is complete, and do not readily explain chromosome organisation at intermediate stages of replication, where newly synthesised DNA remains intermingled~\cite{Harju2024,Brahmachari2026}. To account for this, active processes such as loop extrusion by structural maintenance of chromosomes (SMC) complexes have been proposed to reorganise the genome and promote large-scale ordering~\cite{Harju2024,Brahmachari2026,Alipour2012,Banigan2020}. While these frameworks capture important aspects of chromosome behavior, they treat replication fork advancement as decoupled from the three-dimensional polymer dynamics, and therefore cannot capture the mechanical consequences of ongoing DNA synthesis -- including the replication-driven forces that perturb nucleoid structure -- nor do they reproduce the dynamical fluctuations and stepwise expansion observed experimentally in the nucleoid. 

A key missing ingredient in current biophysical models of the bacterial chromosome is that DNA replication continuously injects new material into the system, driving the nucleoid far from equilibrium while simultaneously modifying its internal connectivity and interaction landscape. In this context, nucleoid-associated proteins (NAPs) mediate effective bridging interactions between DNA segments, providing a natural mechanism for compaction and clustering within the confined cellular volume. At the same time, such interactions introduce effective self-attraction that can compete with chromosome segregation during replication. We hypothesize that the interplay between replication-driven growth and NAP-mediated interactions gives rise to a dynamically reorganising nucleoid, in which clustering, stress buildup, and release collectively shape chromosome dynamics. This competition introduces a nonlinear dependence on interaction strength, leading to qualitatively different organizational regimes. To test this idea, we employ a three-dimensional polymer model (PolyRep)~\cite{Forte2024} that explicitly simulates DNA replication and incorporates protein-mediated interactions.

Using this framework, we show that NAP-mediated interactions give rise to dynamic clustering of DNA, which in turn generates density-wave fluctuations in the nucleoid. When coupled to replication-driven growth, these clusters undergo cycles of buildup and release that produce stepwise expansion dynamics consistent with experimental observations. We find that chromosome segregation emerges naturally in this regime, without the need for dedicated motor-driven mechanisms, but only within a finite window of interaction strengths: weak interactions fail to generate sufficient organization, whereas strong interactions hinder segregation through excessive compaction. Within this optimal regime, replication also promotes the spontaneous formation of replication factories, linking spatial organisation to functional activity. Together, our results demonstrate that the interplay between non-equilibrium growth and protein-mediated interactions provides a unified physical mechanism for nucleoid dynamics and segregation.

\section*{The model}

\begin{figure*}[t!]
    \centering
    \includegraphics[width=1.0\textwidth]{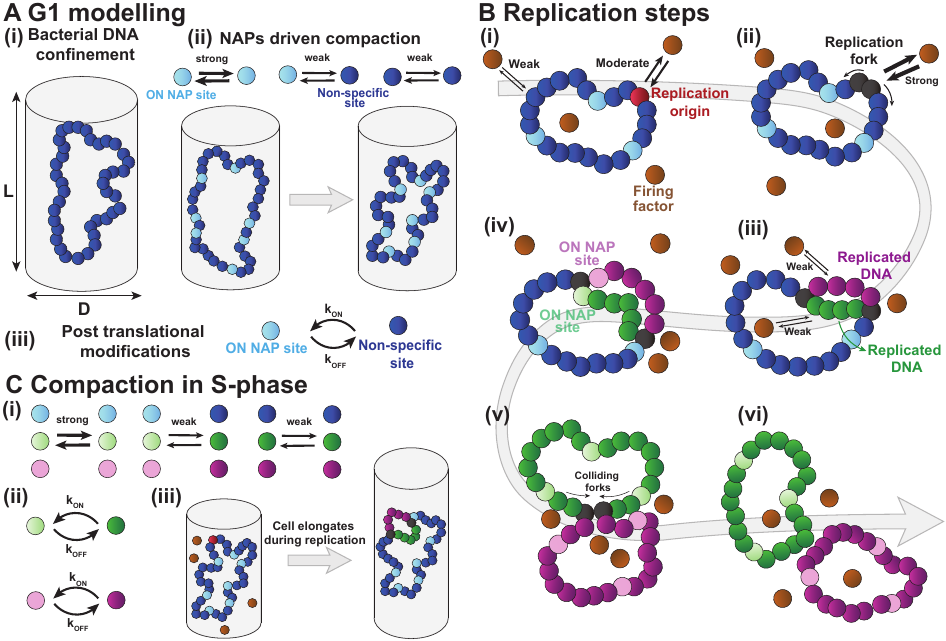}
    \caption{\textbf{A} (i) The bacterial DNA is modelled as a coarse-grained polymer ring confined within a cylindrical volume of diameter $D$ and length $L$. (ii) Nucleoid compaction is driven by NAP-mediated bridging, implemented as effective pairwise interactions. Specific ON NAP binding sites (light blue) interact strongly with each other and weakly with non-specific DNA sites (dark blue), whereas non-specific DNA sites interact weakly among themselves. (iii) NAP post-translational modifications are incorporated by allowing  NAP binding sites to switch between an ON and an OFF state with rates $k_{\mathrm{on}}$ and $k_{\mathrm{off}}$, respectively. OFF NAP binding sites are energetically equivalent to non-specific sites and are therefore visually indistinguishable from them in the figure. The set of beads comprising both ON and OFF NAP binding sites—referred to as potential NAP binding sites—remains unchanged throughout the simulation. \textbf{B} Replication is implemented following the PolyRep framework introduced in Ref.~\cite{Forte2024}. (i) A single DNA bead represents the replication origin (red) and experiences a moderate attraction to multivalent firing factors (FFs). FFs, in turn, weakly attract both non-specific DNA sites and ON NAP binding sites. (ii) When an FF approaches the replication origin within a threshold distance, replication is initiated. The origin and one of its nearest-neighbour beads are converted into two replication forks, which strongly attract FFs and move in opposite directions whenever an FF is nearby. (iii) The progression of each replication fork converts the template DNA into replicated DNA (dark green) and simultaneously generates a newly synthesized daughter strand (dark purple). Both replicated strands interact weakly with FFs, similarly to the template strand. (iv) The subset of beads corresponding to potential NAP binding sites (ON and OFF states) preserves its identity during replication: when a potential NAP binding site is duplicated, both daughter copies remain potential NAP binding sites. In the panel, replicated ON NAP binding sites are shown as light green and light purple beads. (v–vi) Upon collision at the end of replication, the two forks annihilate each other (panel (v)), resulting in the formation of two independent replicated rings (panel (vi)). \textbf{(C)} (i) The template and the two replicated DNA strands are energetically equivalent: all ON NAP binding sites (light blue, green, and purple) interact strongly with each other and weakly with all non-specific sites (dark blue, green, and purple), while non-specific sites interact weakly among themselves. (ii) Replicated potential NAP binding sites undergo post-translational modifications, switching between an ON state and an OFF state (energetically equivalent to non-specific sites) with rates $k_{\mathrm{on}}$ and $k_{\mathrm{off}}$, respectively. (iii) The cylindrical confinement expands linearly with replication progression, thereby mimicking the growth of the bacterial cell.  }
    \label{fig:Fig_model}
\end{figure*}

\subsection*{G1 phase and nucleoid compaction}

To mimic bacterial DNA replication, we constructed a polymer-based model initialized in G1-like conditions. The chromosome is represented as a flexible ring polymer of $N=1000$ beads, such that, if the \textit{E. Coli} genome  ($\sim 4.6 \, Mbp$) was considered, each bead of diameter $\sigma$ would correspond to $\sim 4.6 \, kbp$. The polymer is confined within a cylinder representing a bacterial cell, with height $L$ and diameter $D$ set to a typical aspect ratio $D:L=1:3$~\cite{Tortora2013,Reshes2008} (Fig.~\ref{fig:Fig_model}A(i)). Under these conditions, the polymer occupies $\sim 1 \%$ of the initial cell volume, in line with realistic bacterial genomes~\cite{Fisher2013,Vendeville2011} (see Materials and Methods).

Adjacent beads are connected by harmonic bonds, and the persistence length is set equal to the bead diameter, reflecting the coarse-grained resolution at which DNA rigidity cannot be explicitly captured. To model nucleoid compaction by NAPs, each bead is classified as either a non-specific DNA site or a potential NAP-binding site (for a total of $N_{NAP}$ sites). Potential NAP binding sites switch stochastically back and forth between an active ON state and an inactive OFF state with rates $k_{OFF}$ and $k_{ON}$(Fig. \ref{fig:Fig_model}A(ii)), reflecting post-translational modifications~\cite{Macek2007,Dilweg2018}.
OFF sites behave like non-specific sites; thus, figures display only ON NAP binding sites (cyan) and non-specific beads (blue). ON sites experience strong mutual attractions mimicking high-affinity NAP bridging, whereas all remaining DNA–DNA interactions are governed by weaker attractions to account for non-specific NAP binding~\cite{Nicodemi2009}. At initialization, half of the $N_{NAP}$ sites are ON, and this fraction remains approximately constant by enforcing $k_{ON} = k_{OFF}$.

Nucleoid compaction is controlled by the number of ON NAP binding sites and the interaction strengths among them. We fix $N_{NAP}=83$ (yielding $\sim 41$ ON sites on average) and vary only the ON–ON attraction strength. All DNA interactions are modeled via a truncated and shifted Lennard–Jones potential,
\begin{equation}
V_{LJ/cut} (r_{i,j}) = \left[ V_{LJ}(r_{i,j}) - V_{LJ}(r_c) \right] \Theta(r_c-r_{i,j})\label{Eq:LJ_potential},
\end{equation}
where $r_{i,j}$ is the distance between the $i-th$ and $j-th$ beads, $r_c$ is a cut-off distance and
\begin{equation}
V_{LJ} (r_{i,j}) =  4 \varepsilon \left[ \left(\frac{\sigma}{r_{i,j}} \right)^{12} - \left(\frac{\sigma}{r_{i,j}} \right)^{6} \right], \label{Eq:LJ}
\end{equation}
where $\varepsilon$ is the interaction energy.
Weak, non-specific DNA-DNA attraction is implemented using $r_c=1.5 \, \sigma$ and $\varepsilon=1.1 \, k_BT$ with $k_B$ being the Boltzmann constant and $T$ the room temperature. For ON NAP binding sites, three parameter sets are examined: ($3 \, k_BT, 2\, \sigma$), ($3 \, k_BT, 3\, \sigma$) and ($4 \, k_BT, 2\, \sigma$). These three configurations are hereafter referred to as $NAP_{\mathrm{weak}}$, $NAP_{\mathrm{med}}$, and $NAP_{\mathrm{strong}}$, respectively.

\subsection*{Modeling replication with PolyRep}

DNA replication in S phase is implemented using the PolyRep framework introduced in Ref.~\cite{Forte2024}. The G1-like polymer contains both non-specific and potential NAP binding sites. Two elements are added to initiate replication (Fig.~\ref{fig:Fig_model}B(i)): (i) one bead is designated as the replication origin (red), and (ii) diffusing “firing factors” (FFs, brown spheres) are introduced into the system. In line with Ref.~\cite{Forte2024}, FFs represent multivalent complexes containing the proteins required for initiation and elongation of replication (polymerases, helicases, initiator proteins).
To capture the weak non-specific affinity typical of many DNA-binding proteins—including some bacterial NAPs~\cite{Stracy2021} and eukaryotic transcription factors~\cite{Sheinman2012}—FFs interact weakly with both non-specific and NAP binding sites through the potential in Eq.~\ref{Eq:LJ_potential} with
$(\varepsilon,r_c)=(2 \, k_BT, 1.8 \, \sigma)$. Their affinity for the origin is moderately stronger (i.e. $(6 \, k_BT, 1.8 \, \sigma)$). Once a FF binds the origin, replication begins, and the origin and one adjacent bead convert into two replication forks (black beads in Fig.~\ref{fig:Fig_model}B(ii)). FFs are strongly attracted to these forks (with ($10 \, k_BT, 1.8 \, \sigma$)), and fork progression is enabled whenever a FF is sufficiently close to it. Forks move in opposite directions along the ring (Fig.~\ref{fig:Fig_model}B(iii)): advancement from site $i$ to $i\pm 1$ converts bead $i$ into replicated DNA (green) and creates a complementary replicated bead (purple). After each replication step, a short equilibration run allows beads near the forks to temporarily cross, helping to resolve local entanglements between template and nascent DNA (see Materials and Methods).  The replicated strands are energetically identical to the template; FFs therefore interact with them through the same weak non-specific attraction. Template NAP binding sites—ON or OFF—are duplicated as NAP binding sites in the daughter strands (light green and light purple beads in Fig.~\ref{fig:Fig_model}B(iv) represent replicated ON NAP binding sites). The ON/OFF state itself is not inherited; instead, each replicated NAP binding site randomly adopts an ON or OFF state, reflecting that post-translational NAP modifications may not be robustly maintained across replication. 

Replication ends when the two forks meet and annihilate, producing two fully replicated rings (Fig. \ref{fig:Fig_model}B(v–vi)). Importantly, although multiple and simultaneous rounds of replication can occur in both fast- and slow-growing bacteria~\cite{Trojanowski2017}, here we consider the simpler scenario of a single replication cycle, in which only two replication forks are present.

\subsection*{Nucleoid compaction in S-phase and cell-growth}
Because replicated chromosomes share the same energetic parameters as the template, they compact through the same NAP-mediated interactions throughout the whole replication process. Since multivalent NAPs can bridge sites independent of chromosome identity, all ON NAP binding sites—on both the template and the two daughter chromosomes—experience the same strong attractions, whereas non-specific DNA sites feel only weak interactions (Fig. \ref{fig:Fig_model}C(i)). Replicated NAP binding sites undergo the same stochastic ON/OFF switching as template sites (Fig. \ref{fig:Fig_model}C(ii)).
Finally, to mimic bacterial cell elongation during S phase, the confining cylinder expands along its longitudinal axis $L$ as replication proceeds (Fig.~\ref{fig:Fig_model}C(iii)). Most bacteria are known to elongate exponentially~\cite{Salman2020}, although linear growth has also been reported~\cite{Fisher2013, Willis2017}. We impose here a linear cellular expansion, such that by the end of replication, the cylindrical confinement has doubled in height. 

\section*{Results}

\begin{figure}[htbp]
    \centering
    \includegraphics[width=0.9\columnwidth]{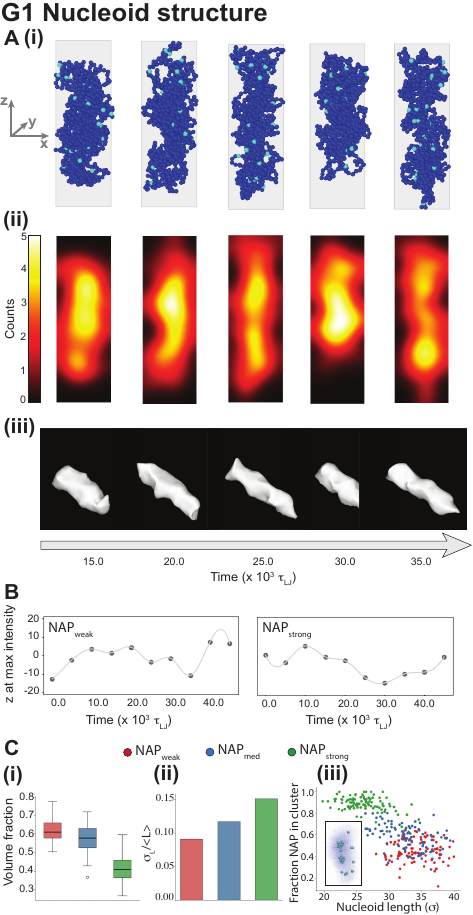}
    {\small
    \caption{ \textbf{A} (i) Time-lapse snapshots of the constrained G1-phase nucleoid prior to replication. Reference axes are indicated on the left. (ii) Corresponding time evolution of nucleoid density for the snapshots shown in panel (i). The colour scale represents the number of DNA beads projected onto the x–z plane, with Gaussian smoothing applied. (iii) Reconstruction of the nucleoid surface for the snapshots in panel (i), illustrating the dynamic nature of the nucleoid structure. \textbf{B} Plots showing the position along the $z$-axis corresponding to the maximum nucleoid density as a function of time. The two panels correspond to independent simulation runs with different values of the attractive energy among NAP binding sites. In both cases, the nucleoid density exhibits fluctuations, but no clear periodicity is observed. \textbf{C} (i) Fraction of the cell volume occupied by  nucleoid in G1 phase for the three different NAP binding-site interaction strengths considered. (ii) Variations in the longitudinal ($z$-axis) nucleoid extension for the three interaction strengths. Notably, although the $NAP_{strong}$ configuration produces the highest nucleoid compaction, it also shows the largest fluctuations in nucleoid extension. (iii) Fraction of ON NAP binding sites engaged in clusters as a function of nucleoid length for the three interaction conditions. The $NAP_{strong}$ case generally forms the largest clusters and the most compact nucleoids; however, the nucleoid extension can increase considerably when these clusters break. Data in panels (i–iii) are obtained from $10$ independent simulations for each NAP binding-site interaction strength.} }
    \label{fig:Fig_G1}
\end{figure}

\subsection*{Dynamic clustering and density fluctuations in the G1 nucleoid}

Before the initiation of replication, bacterial DNA is compacted into a nucleoid-like structure, consistent with structures observed experimentally during the G1 phase. Figure~\ref{fig:Fig_G1}A(i) illustrates representative snapshots of the temporal evolution of a nucleoid in $NAP_{weak}$ conditions. The nucleoid exhibits pronounced dynamical behavior rather than forming a static structure. Specifically, it spans the full radial extent of the confining cylinder, while its longitudinal extension fluctuates over time. These fluctuations in nucleoid length are directly associated with variations in nucleoid density. By projecting the nucleoid onto the XZ plane and calculating its two-dimensional density, spatial density fluctuations along the Z axis become evident (Fig. 2(ii)). The nucleoid undergoes cycles of compaction and extension, and regions of maximum density (highlighted in white and yellow) can transiently split into two distinct domains, as observed in the leftmost panel of Fig.~\ref{fig:Fig_G1}A(ii).
The pronounced rearrangements of the internal nucleoid structure is robust across interaction strengths. By tracking the axial (z) displacement of the density maximum, we observe comparable fluctuations even for the strongest attractive interaction considered in this study $NAP_{strong}$ (Fig. 2B). However, no clear periodicity can be discerned from these temporal fluctuations. The three-dimensional reconstruction of the bacterial nucleoid surface (Fig.~\ref{fig:Fig_G1}A(iii)) further highlights the high degree of structural plasticity of the nucleoid. In contrast to microscopy observations~\cite{Fisher2013,Hadizadeh2012}, however, no helical organization is evident, and quantitative analysis confirms the absence of global chirality (Fig. S1). 

As noted above, fluctuations in both nucleoid length and density are observed across all three parameter sets analysed. Overall, the simulations yield nucleoid volumes ranging from approximately $30\%$ to $80 \%$ of the confining cylindrical cell volume (Fig.~\ref{fig:Fig_G1}C(i)), in agreement with experimentally realistic values. Importantly, the extent of nucleoid length fluctuations depends on the interaction strength between NAP binding sites, with the strongest interaction $NAP_{strong}$ producing the largest variations in longitudinal extension (Fig.~\ref{fig:Fig_G1}C(ii)). These changes are driven by the dynamic formation and dissociation of NAP clusters (i.e. clusters of ON NAP binding sites), which were also observed in STORM experiments~\cite{Wang2011}. In the $NAP_{weak}$ setup, only small clusters form and their sizes remain relatively stable (see Fig. S2). In contrast, the $NAP_{strong}$ scenario promotes the formation of larger clusters, whose ongoing fusion and fission events lead to pronounced changes in nucleoid length. This behavior is further highlighted when plotting the fraction of NAP binding sites within clusters against nucleoid length (Fig.~\ref{fig:Fig_G1}C(iii)). While no clear correlation is observed for the $NAP_{weak}$ condition, the two quantities are strongly anti-correlated in the other two scenarios (Pearson's correlation $r=-0.38$ to $r=-0.5$, $p_{value} < 10^{-4}$). 
Collectively, these results demonstrate that attractive interactions among NAP binding sites give rise to dynamic clustering that drives large-scale nucleoid rearrangements, including density fluctuations and variations in longitudinal extension. This clustering dynamics provides a physical basis for the nucleoid density waves observed experimentally and sets the stage for the non-equilibrium behavior that emerges during replication.
\\

\subsection*{Step-wise nucleoid expansion during replication}

\begin{figure}[!t]
    \centering
    \includegraphics[width=1.0\columnwidth]{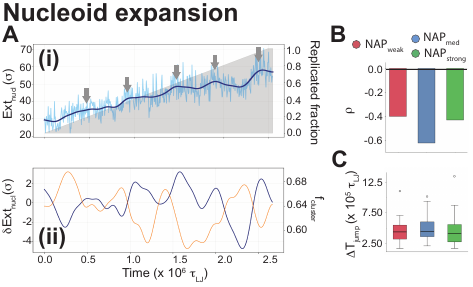}
    \caption{\textbf{A} (i) Representative time evolution of the nucleoid extension  along the $z$-axis, $Ext_{nucl}$ (light and dark blue lines representing raw and smoothed data respectively), for a simulation run with interaction set $NAP_{med}$. The grey region indicates the replicated fraction (right $y$-axis), while grey arrows mark abrupt expansion events superimposed on the overall linear growth. (ii) Deviations of the nucleoid extension from its linear trend, $\delta Ext$ (blue curve, left $y$-axis), together with the fraction of ON NAP binding sites in clusters, $f_{\mathrm{cluster}}$ (orange curve, right $y$-axis), shown as a function of time. \textbf{B} Pearson correlation coefficient between $\delta Ext$ and $f_{\mathrm{cluster}}$ for the three NAP binding-site interaction strengths. \textbf{C} Distributions of the time intervals between consecutive nucleoid extension jumps for the three interaction sets. Panels B and C are computed from $10$ independent simulations for each interaction condition. }
    \label{fig:Fig_expansion}
\end{figure}

During replication, experiments reveal that the nucleoid undergoes a striking stepwise expansion dynamics, in which spells of gradual growth are punctuated by abrupt increases in extension~\cite{Fisher2013}. 
To investigate whether this behavior is reproduced also in our simulations, we measure the extension of the full DNA assembly, namely the composite object formed by the template (blue) and the two daughter (green and purple) nucleoids (see Supplementary Video 1 for an example of a replication run). Since replication pauses are rare in our simulations and the bacterial cell length is imposed to increase linearly with replication progression, one would expect the overall nucleoid extension, $Ext_{nucl}$, to exhibit a comparable linear growth. Consistently, single-trajectory analysis shows that  $Ext_{nucl}$ increases approximately linearly over the course of replication (Fig.~\ref{fig:Fig_expansion}A(i), the grey area representing the replicated fraction). Superimposed on this global linear trend, however, we observe pronounced short-term fluctuations. These are characterized by transient decreases in extension followed by abrupt expansion events; the latter typically occur five or six times per simulation run (grey arrows in Fig.~\ref{fig:Fig_expansion}A(i)). Such abrupt changes in $Ext_{nucl}$, reminiscent of the stepwise expansion experimentally observed, indicate rapid internal structural rearrangements occurring on timescales shorter than that of overall cell growth. 

To elucidate the origin of these fluctuations, we analyze the clustering dynamics of NAP binding sites across both template and daughter strands. We find a strong negative correlation between the fraction of clustered binding sites, $f_{\mathrm{cluster}}$, and deviations from linear nucleoid extension, $\delta Ext_{\mathrm{nucl}}$ (Fig.~\ref{fig:Fig_expansion}A(ii)), indicating that nucleoid contraction is associated with cluster formation, while expansion follows cluster dissolution. This behavior arises from the interplay between NAP-mediated attraction and replication-driven growth. Because the template and the two daughter strands contain a finite number of specific (NAP-binding) sites that experience strong mutual attraction, clusters of NAP-binding sites are expected to self-arrest, leading to finite-size aggregates~\cite{Brackley2016, Chiang2020}. However, replication proceeds concurrently, continuously injecting new DNA (and thus new NAP binding sites) and generating active, out-of-equilibrium perturbations. As a result, clustering becomes intrinsically dynamic: aggregation of NAP binding sites increases effective bridging and drives nucleoid contraction, whereas the appearance of additional binding sites increases cluster size and destabilizes existing aggregates, promoting their rupture and a consequent rapid expansion of the nucleoid. These findings are further supported by simulations in which nucleoid compaction is removed and such abrupt expansions during replication are entirely absent (see Fig.~S3).

The relationship between $\delta Ext_{\mathrm{nucl}}$ and $f_{\mathrm{clusters}}$ is consistently observed across all three simulation conditions considered, corresponding to different strengths of the attractive interaction between NAP binding sites. By pooling the results from $10$ independent simulation runs for each parameter set, we obtain a Pearson correlation coefficient in the range $[-0.63, -0.40]$, with $p$-values smaller than $10^{-10}$ (Fig.~\ref{fig:Fig_expansion}B), confirming a robust negative association between clustering and nucleoid extension fluctuations. 

The time series of $\delta Ext_{\mathrm{nucl}}$ can also be used to extract the time interval between consecutive expansion events, denoted $\Delta T_{\mathrm{jump}}$ (see Supplementary Information for details). Interestingly, the distributions of $\Delta T_{\mathrm{jump}}$ obtained for the three simulation conditions are remarkably similar (Fig.~\ref{fig:Fig_expansion}C). This observation can be rationalised by noting that, although the overall degree of nucleoid compaction and the typical size of NAP-binding-site clusters vary across the three parameter sets, the replication dynamics remain essentially unchanged. In particular, the total replication time is nearly identical in all cases, as pauses are rare. As a result, the temporal statistics of nucleoid expansion events are primarily governed by replication progression rather than by the specific strength of NAP-mediated attractions, leading to comparable jump dynamics across conditions.

Together, these results demonstrate that replication-driven growth and NAP-mediated interactions generate an intrinsic stick–release dynamics in the nucleoid, providing a physical mechanism for the stepwise expansion observed experimentally. This behavior is reminiscent of stress accumulation and intermittent release in driven soft and active materials, where internal activity leads to cycles of loading and structural rearrangement~\cite{Marchetti2013, Natesan2021}. In this context, the bacterial nucleoid can be viewed as an active, non-equilibrium polymer system in which replication acts as an internal drive that continuously perturbs and reorganises its structure.

\subsection*{Stick-release dynamics drives segregation of replicated nucleoids}

\begin{figure}[!t]
    \centering
    \includegraphics[width=1.0\columnwidth]{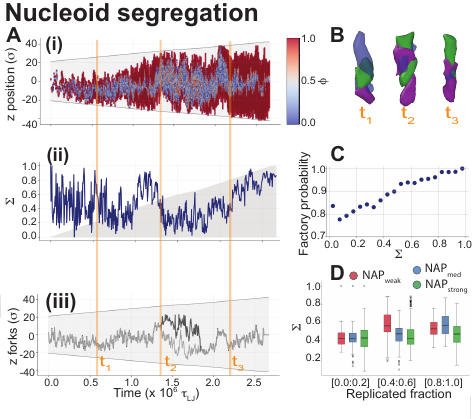}
    \caption{\textbf{A} (i) Kymograph showing the local segregation parameter $\phi$ as a function of $z$ and time for a representative simulation of the $NAP_{med}$ case. (ii) Corresponding time evolution of the global segregation parameter $\Sigma$. The light grey region indicates the replication fraction (refer to the $y$-axis). (iii) Time evolution of the $z$ positions of the two replication forks (light and dark grey curves). The diverging light grey region represents the longitudinal size of the cylindrical cell confinement (refer to the $y$-axis). \textbf{B} Surface reconstructions of the three nucleoid structures—the template (blue) and the two replicated nucleoids (green and purple)—at three different time points corresponding to the simulation shown in panel A. \textbf{C} Replication factory probability versus global segregation parameter $\Sigma$. The factory probability is defined as the probability that the two forks occupy the same NAP binding-site cluster. The plot combines data from all simulations across the three parameter sets.  \textbf{D} Distributions of the global segregation parameter $\Sigma$ at three distinct stages of replication for the three NAP binding-site interaction strengths considered. Each box plot is derived from 10 independent simulation runs at the given interaction strength. }
    \label{fig:Fig_segregation}
\end{figure}

We next ask how the stick–release dynamics identified above controls chromosome segregation during replication. As shown in the previous section, replication-driven growth and NAP-mediated interactions generate intermittent cycles of nucleoid contraction and expansion. These stepwise rearrangements not only shape nucleoid dynamics but also reorganize the spatial distribution of newly synthesized DNA. Starting from a compacted G1 nucleoid, DNA replication progressively generates two daughter filaments that interact through the same NAP-mediated attractions as the template. Under these conditions, the emergence of two spatially segregated nucleoids is not trivial, as inter-filament interactions energetically favor mixing. We therefore investigate how the non-equilibrium dynamics arising from replication and clustering impacts the segregation of replicated chromosomes.

To quantify segregation during replication, we partition the longitudinal ($z$) axis of the cell into discrete bins and define, for each bin, the local segregation parameter
\begin{equation}
\phi(z,t) =
\frac{\left| N_g(z,t) - N_p(z,t) \right|}
{N_g(z,t) + N_p(z,t)} ,
\end{equation}
where $N_g(z,t)$ and $N_p(z,t)$ denote the number of beads belonging to the green and purple daughter filaments, respectively, within a bin centered at position $z$ and at time $t$. By construction,
\begin{equation}
0 \leq \phi(z,t) \leq 1 ,
\end{equation}
with $\phi = 0$ corresponding to perfect mixing ($N_g = N_p$) and $\phi = 1$ to complete local demixing (only one filament present in the bin).
Figure~\ref{fig:Fig_segregation}A(i) shows a kymograph of $\phi(z,t)$ obtained from a single simulation run, together with the linearly increasing cell length (grey region). The parameter $\phi$ approaches unity near the cell poles, indicating spatial segregation of the daughter nucleoids, whereas it decreases near mid-cell, where the two filaments spatially overlap and mix.
To obtain a global measure of segregation at a given time $t$, we define
\begin{equation}
\Sigma(t) =
\frac{N_{\mathrm{mono}}(t)}
{N_{\mathrm{multi}}(t)} ,
\end{equation}
where $N_{\mathrm{mono}}(t)$ is the number of $z$-bins containing exclusively green or purple beads (i.e., $\phi = 1$), and $N_{\mathrm{multi}}(t)$ is the number of bins where both filaments are present (i.e., $\phi < 1$). In this framework, $\Sigma = 1$ indicates complete segregation along the $z$ axis, whereas $\Sigma = 0$ corresponds to complete mixing of the two daughter nucleoids.
The temporal evolution of $\Sigma(t)$ for the same trajectory shown in Fig.~\ref{fig:Fig_segregation}A(i) is reported in Fig.~\ref{fig:Fig_segregation}A(ii). Large values of $\Sigma$ coincide with extended regions where $\phi = 1$, indicating predominant demixing or segregation. Notably, $\Sigma$ exhibits pronounced temporal fluctuations throughout replication. This non-monotonic behavior of $\Sigma(t)$ is consistent with the intermittent expansion dynamics described above, where episodes of nucleoid rearrangement transiently increase mixing (during contraction) before facilitating subsequent spatial separation (upon release). 

More specifically, at time $t_1$, when approximately $20\%$ of the template DNA has been replicated (grey area in Fig.~\ref{fig:Fig_segregation}A(ii)), we find $\Sigma(t_1) \simeq 0.7$, implying that mixing occurs in only about $30\%$ of the $z$-bins. This large segregation is also evident in the corresponding simulation snapshot (Fig.~\ref{fig:Fig_segregation}B, left), where the template (blue) nucleoid is shown for reference.
However, at time $t_2$, at approximately $50\%$ replication, $\Sigma$ markedly decreases, indicating substantial longitudinal intermingling of the daughter nucleoids (Fig.~\ref{fig:Fig_segregation}B, center). Interestingly, this reduction in $\Sigma$ coincides with a disruption of the replication factory formed by the two replication forks. This event is visible as a splitting of the fork trajectories in the $z$-versus-time representation (Fig.~\ref{fig:Fig_segregation}A(iii)), suggesting a coupling between replication-factory integrity and large-scale nucleoid segregation dynamics. 

A deeper understanding of the role of the replication factory in promoting segregation can be achieved by pooling together all simulation datasets obtained for the three different values of the attractive energy between NAP binding sites, and investigating the relationship between the global segregation parameter $\Sigma$ and  the probability of observing a replication factory -- defined as the condition in which the two replication forks remain in close spatial proximity in $3D$. The resulting dependence is shown in Fig.~\ref{fig:Fig_segregation}C. A clear positive correlation emerges between factory probability and global segregation, indicating that an increased likelihood of fork colocalisation corresponds to higher values of $\Sigma$, and thus to enhanced spatial segregation of the replicated nucleoids. Notably, the probability of observing a replication factory remains consistently high ($>0.7$). This behaviour differs from what was reported in our previous work~\cite{Forte2024}, where replication factories were not observed when very weak non-specific attraction between FFs and the chromatin filament — comparable in magnitude to the FF–DNA attraction used here — was considered. The discrepancy highlights the crucial role of bacterial confinement and of NAP-mediated clustering in stabilising fork colocalisation. In the present model, these effects are sufficient to maintain the two forks spatially proximate even when FFs experience attractive interactions exclusively with the forks and steric repulsion with all other components of the system (see Fig.S4), a result which was not observed in Ref.~\cite{Forte2024}. Importantly, we observe that the two replication forks naturally localize near mid-cell (Fig.~\ref{fig:Fig_segregation}A(iii)), driven by the replication and segregation dynamics themselves. 

Following the drop in global segregation at mid-replication, the system investigated in Fig.~\ref{fig:Fig_segregation}A regains segregation once the two replication forks come into close spatial proximity. At time $t_3$, when approximately $80\%$ of the template DNA strand has been replicated, the segregation parameter $\Sigma$ increases again to larger values, indicating renewed spatial separation of the two nucleoids (Fig.~\ref{fig:Fig_segregation}B, right). The non-monotonic behavior of $\Sigma(t)$ across replication stages is consistently observed in all three simulation set-ups considered (Fig.~\ref{fig:Fig_segregation}D). During the early stages of replication (less than $20\%$ replicated DNA), attraction among NAP-binding sites does not significantly influence global segregation: all three cases display similar values of $\Sigma$, exceeding $0.4$. Visual inspection of simulation configurations suggests that $\Sigma \approx 0.4$ represents a threshold above which the two nucleoids appear as demixed entities. At mid-replication ($40\%$--$60\%$ replicated DNA), the systems with the strongest NAP-mediated attraction ($NAP_{med}$ and $NAP_{strong}$) exhibit the greatest degree of mixing between the replicated nucleoids. However, segregation increases again toward the end of replication, with the median $\Sigma$ exceeding $0.4$ in all three setups. This transient decrease followed by recovery of segregation during replication is consistent with previous studies in \textit{B.~subtilis}~\cite{El2020} and \textit{E.~coli}~\cite{Gelber2021}, which reported stochastic patterns in the bulk segregation of replicated nucleoids. 

Our simulations therefore demonstrate that 
segregation is driven by the interplay between compaction and replication-driven dynamics. This is because the intermittent stick–release behavior of the nucleoid generates repeated cycles of mixing and separation, which progressively enhance the spatial demixing of daughter chromosomes. Consistent with this picture, segregation is significantly weaker in the absence of NAP-mediated clustering (Fig.~S5), where both stepwise expansion dynamics and large-scale nucleoid rearrangements are strongly suppressed. Together, these results show that chromosome segregation emerges as a direct consequence of non-equilibrium nucleoid dynamics driven by replication and modulated by protein-mediated interactions.

\if{
\subsection*{Simulations show a step-wise expansion of the nucleoid during replication}
\begin{figure}[!t]
    \centering
    \includegraphics[width=1.0\columnwidth]{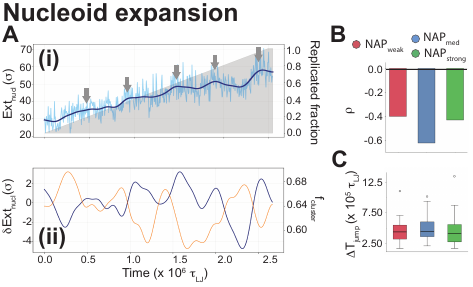}
    \caption{\textbf{A} (i) Representative time evolution of the nucleoid extension along the $z$-axis for a simulation run with interaction set $NAP_{med}$. The grey region indicates the replicated fraction (right $y$-axis), while grey arrows mark abrupt expansion events superimposed on the overall linear growth. (ii) Deviations of the nucleoid extension from its linear trend, $\delta Ext$ (blue curve, left $y$-axis), together with the fraction of NAP binding sites in clusters, $f_{\mathrm{cluster}}$ (orange curve, right $y$-axis), shown as a function of time. \textbf{B} Pearson correlation coefficient between $\delta Ext$ and $f_{\mathrm{cluster}}$ for the three NAP binding-site interaction strengths. \textbf{C} Distributions of the time intervals between consecutive nucleoid extension jumps for the three interaction sets. Panels B and C are computed from $10$ independent simulations for each interaction condition. }
    \label{fig:Fig_expansion}
\end{figure}

We next investigate how the nucleoid expands along the $z$-axis during replication, in line with microscopy measurements~\cite{Fisher2013}.  Because conventional fluorescence microscopy does not allow discrimination between the template and replicated nucleoids, we quantify here the $z$ extension of the entire DNA assembly, namely the composite object formed by the template (blue) and the two daughter (green and purple) nucleoids shown in Fig.~\ref{fig:Fig_segregation}B. Since replication pauses are rare in our simulations and the bacterial cell length is imposed to increase linearly with replication progression, one would expect the overall nucleoid extension, $Ext_{nucl}$, to exhibit a comparable linear growth. Consistently, single-trajectory analysis shows that the $Ext_{nucl}$ increases approximately linearly over the course of replication (Fig.~\ref{fig:Fig_expansion}A(i), the grey area representing the replicated fraction). Superimposed on this global linear trend, however, we observe pronounced short-term fluctuations. These are characterized by transient decreases in extension followed by abrupt expansion events (grey arrows in Fig.~\ref{fig:Fig_expansion}A(i)). Such abrupt changes in $Ext_{nucl}$ indicate rapid internal structural rearrangements occurring on timescales shorter than that of overall cell growth. To elucidate the origin of these fluctuations, we performed a clustering analysis on NAP binding sites, considering NAP binding sites belonging to both the template and daughter strands, as they are energetically equivalent. This analysis reveals a negative correlation between the fraction of NAP binding sites engaged in clusters, $f_{cluster}$, and the deviation of the nucleoid extension from its linear growth trend, $\delta Ext_{nucl}$ (Fig.~\ref{fig:Fig_expansion}A(ii)), suggesting that contractions and jump-like increases in $Ext_{nucl}$ are linked to the dynamic clustering of NAP binding sites. Because the three DNA filaments (the template and the two daughter strands) contain a finite number of specific (NAP-binding) sites that experience strong mutual attraction, clusters of NAP-binding sites are expected to self-arrest, leading to finite-size aggregates~\cite{Brackley2016, Chiang2020}. However, replication proceeds concurrently, continuously injecting new DNA (and thus new NAP binding sites) and generating active, out-of-equilibrium perturbations. As a result, clustering becomes intrinsically dynamic: aggregation of NAP binding sites increases effective bridging and drives nucleoid contraction, whereas the appearance of additional binding sites increases cluster size and destabilizes existing aggregates, promoting their rupture and a consequent rapid expansion of the nucleoid. Importantly, a qualitatively similar dynamical behaviour of the total nucleoid extension was reported experimentally in \textit{E. Coli}~\cite{Fisher2013}. In that study, the nucleoid length increased overall during growth but exhibited approximately four abrupt expansion events, each preceded by transient reductions in extension. These discontinuities coincided with previously identified nucleoid transitions associated with increased sister-chromosome separation and were interpreted as reflecting the sequential release of chromosomal tethers. \\
The relationship between $\delta Ext_{\mathrm{nucl}}$ and $f_{\mathrm{clusters}}$ is consistently observed across all three simulation conditions considered, corresponding to different strengths of the attractive interaction between NAP binding sites. By pooling the results from $10$ independent simulation runs for each parameter set, we obtain a Pearson correlation coefficient in the range $[-0.63, -0.40]$, with $p$-values smaller than $10^{-10}$ (Fig.~\ref{fig:Fig_expansion}B), confirming a robust negative association between clustering and nucleoid extension fluctuations. \\
The time series of $\delta Ext_{\mathrm{nucl}}$ can also be used to extract the time interval between consecutive expansion events, denoted $\Delta T_{\mathrm{jump}}$ (see Supplementary Information for details). Interestingly, the distributions of $\Delta T_{\mathrm{jump}}$ obtained for the three simulation conditions are remarkably similar (Fig.~\ref{fig:Fig_expansion}C). This observation can be rationalised by noting that, although the overall degree of nucleoid compaction and the typical size of NAP-binding-site clusters vary across the three parameter sets, the replication dynamics remain essentially unchanged. In particular, the total replication time is nearly identical in all cases, as pauses are rare. As a result, the temporal statistics of nucleoid expansion events are primarily governed by replication progression rather than by the specific strength of NAP-mediated attractions, leading to comparable jump dynamics across conditions.
}\fi

\section*{Discussion}

In summary, we have shown that the bacterial nucleoid behaves as an active, non-equilibrium polymer in which DNA replication and protein-mediated interactions jointly determine its spatiotemporal organisation. Using the PolyRep framework~\cite{Forte2024}, which explicitly models DNA synthesis, we identify a unifying physical mechanism whereby nucleoid-associated protein (NAP)-mediated clustering, coupled to replication-driven growth, generates intermittent stick–release dynamics. This mechanism underlies a hierarchy of phenomena observed in our simulations, including density fluctuations in the G1 phase, stepwise nucleoid expansion during replication, and the emergence of chromosome segregation.

The main result of our study is that diverse dynamical features of the nucleoid arise from a common physical origin: the clustering dynamics of NAP binding sites. In the G1 phase, stochastic formation and dissolution of clusters generate pronounced density fluctuations, giving rise to wave-like rearrangements of nucleoid structure (Fig.~\ref{fig:Fig_G1}). During replication, the same clustering mechanism is driven away from equilibrium by the continuous injection of newly synthesised DNA. This process destabilises existing clusters, leading to cycles of stress buildup and release that manifest as stepwise nucleoid expansion (Fig.~\ref{fig:Fig_expansion}). The robustness of these dynamics across interaction strengths reflects the fact that their characteristic timescales are set primarily by replication progression rather than by equilibrium clustering properties. 
The average frequency of expansion events is consistent with experimental observations, supporting the idea that these dynamics are governed (at least to a first approximation) by the competition between nonlinear NAP-mediated attraction and replication-driven growth.

A key consequence of the non-equilibrium stick–release dynamics is the emergence of chromosome segregation (Fig.~\ref{fig:Fig_segregation}). In our simulations, segregation arises from repeated cycles of nucleoid contraction and expansion that alternately promote mixing and spatial separation of replicated DNA. 
This stick-release behaviour leads to a characteristic non-monotonic evolution of the global segregation parameter, with a transient reduction at mid-replication followed by recovery, consistent with the stochastic segregation patterns observed in both \textit{B. subtilis} and \textit{E. coli}~\cite{El2020,Gelber2021}. 

Mechanistically, stepwise expansion corresponds to the release of accumulated stress, in agreement with the hypothesis of Ref.~\cite{Fisher2013}. Stress builds up through the interplay between NAP-mediated compaction, which promotes local crowding and entanglement, and replication-driven forces, which increases the energetic cost of these constrained configurations. The resulting instability leads to abrupt structural rearrangements that release stored stress, promoting segregation. Consistent with this picture, segregation is strongly reduced in the absence of NAP-mediated clustering (Fig.~S5), where stress accumulation and release are suppressed (Fig~S3). In this regime, replication proceeds more linearly, but entanglements persist for longer as they are no longer eliminated by the stick-release mechanism, leading to reduced segregation efficiency. These findings demonstrate that chromosome segregation emerges as a dynamical balance that is fundamentally dependent on the non-equilibrium nature of the bacterial chromosome.

Another emergent feature of this dynamically reorganising nucleoid is the formation of replication factories (Fig.~\ref{fig:Fig_segregation}) -- defined by the colocalisation of the two replication forks and observed also in eukaryotes~\cite{Dasaro2024}. These form due to NAP-mediated clustering and cellular confinement, with the two forks localising near mid-cell as a consequence of the replication and segregation dynamics themselves, rather than through any externally imposed constraint. The probability of observing a replication factory correlates positively with the global segregation parameter, suggesting that factory integrity plays a functional role in promoting large-scale nucleoid segregation.

Our results place the bacterial nucleoid within the class of active soft matter systems~\cite{Marchetti2013}, in which internal driving continuously perturbs structure and leads to intermittent, collective rearrangements. Here, DNA replication acts as a source of activity that injects material and stress into the system, while NAP-mediated interactions provide an effective cohesive force. The resulting interplay gives rise to behaviour reminiscent of driven soft materials, where stress accumulation and release govern large-scale dynamics.
Several aspects of bacterial chromosome biology are not captured by the present model and represent natural directions for future work. Our simulations do not reproduce the helical organisation of the nucleoid observed in fluorescence microscopy experiments~\cite{Fisher2013}. A key missing ingredient is likely the organisation of bacterial DNA into supercoiled loops stabilised by SMC complexes, which imparts a bottle-brush architecture to the chromosome. Previous work on bottle-brush polymers confined in cylindrical geometries has shown that their backbone naturally adopts a helical arrangement~\cite{Chaudhuri2012, Jung2019}, suggesting that incorporating loop extrusion and supercoiling into PolyRep might recover this feature. 
We have also restricted our analysis to a single replication cycle, whereas fast-growing bacteria undergo multifork replication~\cite{Wang2009}; extending PolyRep to this regime will be important for understanding chromosome organisation under rapid growth conditions. \\
Finally, we have not considered the effects of macromolecular crowding, which has been shown to influence both nucleoid compaction~\cite{Yang2020} and sister chromatid segregation~\cite{Chang2025}. Incorporating these additional ingredients in future work will allow for a more complete understanding of how the different physical and biochemical components of the bacterial cell interact to organise the genome and ultimately enable a fully quantitative description of bacterial chromosome dynamics.


\section{Materials and Methods}

\subsection*{Bacterial DNA in G1} DNA is modeled as a coarse-grained polymer consisting of $1000$ beads, each with diameter $\sigma$. The mapping between $\sigma$ and the number of DNA base pairs depends on the specific bacterial genome considered. For example, when modeling \textit{E. coli}, whose genome length is approximately $4.6 , \mathrm{Mbp}$, each bead represents about $5 \, \mathrm{kbp}$. \\
After the formation of the initial configuration and a short equilibration (see SI for more details), the energy potentials of the system are as follows. Adjacent beads are connected by harmonic springs described by the potential
\begin{equation}
V_H(r) = K_H (r - R_H)^2, \label{Eq:harmonic_methods}
\end{equation}
where $R_H = 1.1 \, \sigma$ is the equilibrium bond length and $K_H = 200 \, k_B T$ is the spring constant. Here, $k_B$ denotes the Boltzmann constant and $T = 300 \, \mathrm{K}$ is the system's temperature.
Polymer bending rigidity is introduced through a cosine bending potential
\begin{equation}
V_B(\phi) = K_B (1 + \cos \phi), \label{Eq:cosine_methods}
\end{equation}
where $\phi$ is the angle formed by three consecutive beads and $K_B$ is the bending rigidity coefficient. Since the effective bead size is substantially larger than the intrinsic DNA persistence length, we set $K_B = 1 \, k_B T$. By comparison with the Kratky–Porod model, this choice corresponds to a persistence length $l_p = 1 \, \sigma$. Under these conditions, the modeled DNA behaves as a fully flexible polymer.
Non-bonded interactions are modeled by a cut and shifted Lennard-Jones potential 
\begin{equation}
    V_{LJ/cut}(r) = [V_{LJ}(r)-V_{LJ}(r_c)]\Theta(r_c-r) \label{Eq:LJ_cut_methods}
\end{equation}
where $r_c$ is the cutoff distance and 
\begin{equation}
V_{LJ} (r) =  4 \varepsilon \left[ \left(\frac{\sigma}{r} \right)^{12} - \left(\frac{\sigma}{r} \right)^{6} \right]. \label{Eq:LJ_methods}
\end{equation}
The interaction energy $\varepsilon$ and the cutoff distance $r_c$ depend on the types of beads involved in the interaction. Before the onset of replication, three types of polymer beads are present: non-specific DNA beads (dark-blue beads in Fig.~\ref{fig:Fig_model}) and nucleoid-associated protein (NAP) binding sites in either the ON or OFF state (light-blue and dark-blue beads in Fig.~\ref{fig:Fig_model}, respectively). Only a predefined subset of DNA beads can act as potential NAP binding sites, and these sites stochastically switch between their ON and OFF states to mimic post-translational modifications. In the OFF state, NAP binding sites behave identically to non-specific DNA beads; consequently, they are not visually distinguishable from non-specific sites in Fig.~\ref{fig:Fig_model}, where both appear as dark-blue beads. Further details regarding the switching mechanism are provided in the next section. \\
Non-specific DNA beads and OFF NAP binding sites interact with all other polymer beads through weak attractive interactions characterized by $\varepsilon = 1.1 \, k_B T$ and $r_c = 1.5 \, \sigma$. In contrast, ON NAP binding sites interact with each other via a stronger attractive interaction, which represents the primary driving force responsible for the compaction of the DNA ring into a nucleoid-like structure. Three sets of parameters $(\varepsilon, r_c)$ are considered in the simulations: $(3 \, k_B T, 2 \, \sigma)$, $(3 \, k_B T, 3 \, \sigma)$, and $(4 \, k_B T, 2 \, \sigma)$. Each parameter set produces a different nucleoid volume fraction, with average values of $0.62$, $0.58$, and $0.41$, respectively. Larger values of the non-specific DNA–DNA attraction would lead to an unrealistically compact nucleoid. Conversely, weaker non-specific interactions would reduce the overall compaction, requiring stronger attractions among NAP binding sites to recover realistic nucleoid densities. Although other combinations of non-specific and NAP-mediated interactions could in principle be explored, we focus here on these parameter sets in order to reproduce representative and biologically plausible levels of nucleoid compaction.\\
The dynamics of a bead located at position $\textbf{r}_i$ are governed by the Langevin equation
\begin{equation}
m_i \frac{\partial^2 \textbf{r}_i}{\partial t^2} = \nabla_i U - \gamma_i \frac{\partial \textbf{r}_i}{\partial t} + \sqrt{2 k_B T \gamma_i},\boldsymbol{\eta}_i , \label{Eq:langevin}
\end{equation}
where $U$ is the total potential energy of the system, $\gamma_i$ is the friction coefficient experienced by the $i$-th bead due to the solvent, and $\boldsymbol{\eta}_i$ represents a Gaussian thermal noise. The components of the noise satisfy $\langle \eta_{i\alpha} \rangle = 0$ and $\langle \eta_{i \alpha}(t)\eta_{j \beta}(t')\rangle = \delta_{ij} \delta_{\alpha \beta} \delta(t-t') \rangle$, where $\delta_{ij}$ and $\delta_{\alpha\beta}$ denote Kronecker deltas and $\delta(t-t')$ is the Dirac delta function.
Simulations are performed using the software \textit{LAMMPS}~\cite{Plimpton1995} with a time step $dt = 0.01$.

\subsection*{Bacterial confinement and post-translational modifications}

The DNA polymer is initially confined within a cylindrical container of diameter $D = 14 \, \sigma$ and length $L = 42 \, \sigma$, corresponding to an aspect ratio $D:L = 1:3$. \\
To account for post-translational modifications, NAP binding sites switch stochastically between an ON and an OFF state with rates $k_{ON}= k_{OFF} \coloneqq k_{switch}$. We set $k_{switch} = 2 \cdot 10^{-5} \, \tau_{LJ}^{-1}$, where $\tau_{LJ}$ denotes simulation time units. The choice of this switching rate can be rationalized by mapping the simulation time scale to that of \textit{E. coli} DNA replication. Additional details about the replication process are provided in the next section; however, in our simulations, the replication of the $1000$-bead polymer is completed in approximately $2.5 \cdot 10^6 \, \tau_{LJ}$. In \textit{E. coli}, chromosome replication typically takes about $\sim 40 \, \mathrm{min}$ (Ref.~\cite{Cooper1968}), which yields a conversion factor $\tau_{LJ} \sim 10^{-3} \, \mathrm{s}$. With this mapping, the switching rate $k_{switch} = 2 \cdot 10^{-5} \, \tau_{LJ}^{-1}$ corresponds to a characteristic switching time of approximately $1 \, \mathrm{min}$. This timescale is consistent with experimental observations reporting rapid post-translational modifications in bacteria and characteristic timescales of seconds to minutes in eukaryotic systems~\cite{Macek2019, Leutert2021}.

\subsection*{Mimicking DNA replication.} 
To mimic DNA replication, a specific DNA bead 
is defined as the replication origin (red bead in Fig.~\ref{fig:Fig_model}). Additional particles are introduced into the system to represent firing factors (FFs, brown beads), representing the molecular components required for replication initiation and elongation. Excluded-volume interactions between FFs are modeled using a Weeks–Chandler–Andersen potential
\begin{equation}
V_{WCA}(r) = 4 k_BT \left[ \left(\frac{\sigma}{r} \right)^{12} - \left(\frac{\sigma}{r} \right)^{6} +\frac{1}{4} \right]\Theta(2^{1/6}\sigma -r).
\end{equation}
Interactions between FFs and DNA beads are described by the cut and shifted Lennard–Jones potential defined in Eq.~\ref{Eq:LJ_cut_methods}. For interactions between FFs and the replication origin, we set $r_c = 1.8 \, \sigma$ and $\varepsilon = 6 \, k_BT$, while interactions between FFs and any other DNA bead (both non-specific DNA sites and NAP binding sites) are characterized by $\varepsilon = 2 \, k_BT$. \\
If a FF approaches the replication origin within a distance $r_{threshold} = 1.8 \, \sigma$, the origin fires with probability $P_{firing}$. In this work, unlike in Ref.~\cite{Forte2024}, we set $P_{firing} = 1$. Smaller values would simply delay the onset of replication, whereas our focus is on the dynamical processes occurring during replication.
Once the origin fires, two replication forks are created (black beads in Fig.~\ref{fig:Fig_model}). One fork is located at the bead that previously represented the origin, while the second is placed on one of its two nearest neighbors. Forks interact with FFs through the Lennard–Jones potential in Eq.~\ref{Eq:LJ_cut_methods} with $\varepsilon = 10 \, k_BT$ and $r_c=1.8 \, \sigma$. When an FF approaches within $r_{threshold}$, forks advance independently in opposite directions along the polymer. The strong affinity between forks and FFs, combined with the compact nucleoid structure of the DNA polymer, results in FFs remaining almost constantly in contact with the forks. As a consequence, fork progression rarely pauses and replication proceeds nearly continuously. \\
Whenever a fork advances, for example, from bead $i$ to bead $i+1$, bead $i$ becomes replicated DNA and a new bead is inserted into the system (green and purple beads in Fig.~\ref{fig:Fig_model}). The two replicated polymers (purple and green strands) are energetically equivalent to the template strand. Consecutive beads within the replicated strands are connected through the harmonic potential defined in Eq.~\ref{Eq:harmonic_methods}, and their bending rigidity is described by Eq.~\ref{Eq:cosine_methods}.
During replication, the newly synthesized polymer (purple strand) is temporarily connected to the two forks through harmonic bonds described by Eq.~\ref{Eq:harmonic_methods}, linking each fork to the corresponding extremity of the growing filament. FFs interact with the replicated polymers in the same way as with the template strand, i.e., through the potential in Eq.~\ref{Eq:LJ_cut_methods} with $r_c = 1.8 \, \sigma$ and $\varepsilon = 2 \, k_BT$. \\
During replication, 
template NAP binding sites are replicated into corresponding NAP binding sites on the two daughter polymers. However, their ON/OFF state is not necessarily conserved. For example, a template ON NAP binding site may be replicated into either an ON or an OFF NAP binding site on the daughter strands with equal probability. This choice accounts for possible post-translational modifications occurring during replication. Replicated NAP binding sites continue to undergo stochastic switching between ON and OFF states with rate $k_{switch} = 2 \cdot 10^{-5} \, \tau_{LJ}^{-1}$, allowing their state to change during the course of replication.\\
As for non-bonded interactions, all DNA beads in the system -- both template and replicated -- are treated equivalently. Non-specific DNA beads (blue, green, and purple beads in Fig.~\ref{fig:Fig_model}) experience weak attractive interactions with all other DNA beads, described by the potential in Eq.~\ref{Eq:LJ_cut_methods} with $r_c = 1.5 \, \sigma$ and $\varepsilon = 1.1 \, k_BT$. In contrast, all ON NAP binding sites (light blue, light green, and light purple beads in Fig.~\ref{fig:Fig_model}) interact strongly with each other, with parameter sets $(\varepsilon, r_c) = (3 \, k_BT, 2 \, \sigma)$, $(3 \, k_BT, 3 \, \sigma)$, and $(4 \, k_BT, 2 \, \sigma)$. \\
The replication process terminates when the two 
forks 
meet 
and annihilate, leaving two 
replicated DNA rings. \\
Throughout replication, the dynamics of the system evolve according to 
Eq.~\ref{Eq:langevin}. In addition, to account for the natural growth of bacterial cells during the S-phase, the cylindrical confinement expands linearly along its longitudinal axis as replication progresses. Specifically, at a given stage of replication, the cylinder length is given by $L^* = L(1+f)$, where $L = 42 \, \sigma$ is the initial length and $f$ is the replication fraction. After replication, when $f=1$, the cylinder length has doubled. Additional details about short equilibration stages during replication are provided in SI.

\section{Acknowledgements}
We acknowledge Stephan Gruber and Sara Buonomo for useful discussions. G.F. acknowledges support from the Leverhulme Trust, United Kingdom (Early Career Fellowship ECF-2024-221). D.M. and G.F. acknowledge support from the Wellcome Trust, United Kingdom (223097/Z/21/Z).

\bibliography{biblio}

\end{document}